\documentclass[%
reprint,
superscriptaddress,
 amsmath,amssymb,
 aps,
prl,
]{revtex4-2}

\usepackage{graphicx}
\usepackage{graphicx}
\usepackage{dcolumn}
\usepackage{float}

\usepackage{subfigure}

\usepackage{amsmath}

\usepackage{booktabs}
\usepackage{bm}

\begin{document}

\preprint{APS/123-QED}

\title{Classification and Generation of Light Sources Using Gamma Fitting}

\author{Shuanghao Zhang}
\affiliation{Electronic Materials Research Laboratory, Key Laboratory of the Ministry of Education $\&$ International Center for Dielectric Research, School of Electronic Science and Engineering, Xi'an Jiaotong University, Xi'an, Shaanxi 710049, China}
\author{Huaibin Zheng}%
\email{huaibinzheng@xjtu.edu.cn}
\affiliation{Electronic Materials Research Laboratory, Key Laboratory of the Ministry of Education $\&$ International Center for Dielectric Research, School of Electronic Science and Engineering, Xi'an Jiaotong University, Xi'an, Shaanxi 710049, China}
\author{Gao Wang}
\affiliation{School of Physics $\&$ Astronomy, University of Glasgow, Glasgow G12 8QQ, UK}%
\author{Hui Chen}
\affiliation{Electronic Materials Research Laboratory, Key Laboratory of the Ministry of Education $\&$ International Center for Dielectric Research, School of Electronic Science and Engineering, Xi'an Jiaotong University, Xi'an, Shaanxi 710049, China}%
\author{Yuchen He}
\affiliation{Electronic Materials Research Laboratory, Key Laboratory of the Ministry of Education $\&$ International Center for Dielectric Research, School of Electronic Science and Engineering, Xi'an Jiaotong University, Xi'an, Shaanxi 710049, China}%
\author{Sheng Luo}
\affiliation{Electronic Materials Research Laboratory, Key Laboratory of the Ministry of Education $\&$ International Center for Dielectric Research, School of Electronic Science and Engineering, Xi'an Jiaotong University, Xi'an, Shaanxi 710049, China}%
\author{Jianbin Liu}
\affiliation{Electronic Materials Research Laboratory, Key Laboratory of the Ministry of Education $\&$ International Center for Dielectric Research, School of Electronic Science and Engineering, Xi'an Jiaotong University, Xi'an, Shaanxi 710049, China}%
\author{Yu Zhou}
\affiliation{MOE Key Laboratory for Nonequilibrium Synthesis and Modulation of Condensed Matter, Department of Applied Physics, Xi'an Jiaotong University, Xi'an 710049, China}%
\author{Zhuo Xu}
\affiliation{Electronic Materials Research Laboratory, Key Laboratory of the Ministry of Education $\&$ International Center for Dielectric Research, School of Electronic Science and Engineering, Xi'an Jiaotong University, Xi'an, Shaanxi 710049, China}%

\date{\today}

\begin{abstract}
In general, the typical approach to discriminate antibunching, bunching or superbunching categories make use of calculating the second-order coherence function ${g^{(2)}}(\tau )$ of light. Although the classical light sources correspond to the specific degree of second-order coherence ${g^{(2)}}(0)$, it does not alone constitute a distinguishable metric to characterize and determine light sources. Here we propose a new mechanism to directly classify and generate antibunching, bunching or superbunching categories of light, as well as the classical light sources such as thermal and coherent light, by Gamma fitting according to only one characteristic parameter $\alpha$ or $\beta$. Experimental verification of beams from four-wave mixing process is in agreement with the presented mechanism, and the influence of temperature $T$ and laser detuning $\Delta$ on the measured results are investigated. The proposal demonstrates the potential of classifying and identifying light with different nature, and the most importantly, provides a convenient and simple method to generate light sources meeting various application requirements according to the presented rules. Most notably, the bunching and superbunching are distinguishable in super-Poissonian statistics using our mechanism.

\end{abstract}

\maketitle

The correlation of two optical intensities, generally expressed in terms of the degree of second-order coherence, was first measured by Hanbury Brown and Twiss \cite{Hanbury1956Correlation,Brown1956A}. Such measurements have a particular significance for the correspondence between the statistics and interference nature of light. The correlation and coherent properties of the light are usually studied in terms of correlation functions of the amplitudes of the field and intensity. In particular, the conventional approaches to discriminate antibunching, bunching or superbunching make use of calculating the second-order coherence function ${g^{(2)}}(\tau )$ of light intensity \cite{Loudon1983The}. The peak value ${{\rm{g}}^{(2)}}(0)$, or the degree of second-order coherence, is typically used to classify light sources as, ${{\rm{g}}^{(2)}}(0) < 1$ referring to photon antibunching \cite{mandel1995optical}, $1 < {{\rm{g}}^{(2)}}(0) < 2$ to photon bunching \cite{2018Highly}, and ${{\rm{g}}^{(2)}}(0) > 2$ to photon superbunching or extrabunching \cite{Ficek2004Quantum}. On the other hand, antibunching corresponds to sub-Poissonian statistics in single mode \cite{1984Quantum}, bunching and superbunching to super-Poissonian statistics \cite{Loudon1983The}. Unfortunately, the bunching and superbunching are indistinguishable in super-Poissonian statistics.

Recent advances in the realization and application of these different categories show the activity and maturity of the research domain. The antibunching effect of the light radiated by the driven atom occurred in nonclassical phenomenons \cite{Ryou2018Strong,2020Light}. The bunching and superbunching effect were observed in linear optical system and nonlinear process \cite{hong2012two,zhou2017superbunching,spasibko2017multiphoton,Allevi2017Super}. In addition, Bromberg \cite{bromberg2010hanbury} and Zhang \cite{Zhang:20} showed the possibility of controlling the bunching properties of thermal light through nonlinear optical effect. Especially in rogue waves \cite{manceau2019indefinite}, the superbunching was demonstrated in extreme heavy-tailed distribution. But the "heavy tail" behavior of light, after squeezing and optical harmonics, must be illustrated by comparing to the incident statistical distribution with the same average intensity of output light, rather than real input light, then calculated second-order correlation function for observation.

For light source identification, typical classical light sources correspond to the specific degree of second-order coherence ${g^{(2)}}(0)$. As is known that ${g^{(2)}}(0)$ equals 2 for thermal light and 1 for coherent light \cite{mandel1995optical}. Conversely, the ${g^{(2)}}(0)$ does not alone constitute a distinguishable metric to characterize and determine light sources, yet there is a confusing fact that the light sources with diverse statistical distribution can obtain the same degree of second-order coherence ${g^{(2)}}(0)$. For example, the exponential distribution and binomial distribution can achieve simultaneously ${g^{(2)}}(0) = 2$ by adjusting different mean and variance values. Moreover, more recent work in light source identification using machine learning were reported \cite{doi:10.1063/1.5133846}, which however skipped the essential and underlying physical rules.

In this letter, we propose a new mechanism to directly classify and generate antibunching, bunching or superbunching categories of light, as well as the typical light sources such as thermal and coherent light, by Gamma fitting according to only one characteristic parameter $\alpha$ or $\beta$. Experimental verification of beams from four-wave mixing process is in agreement with the presented mechanism, and the influence of temperature $T$ and laser detuning $\Delta$ on the measured results are investigated. In contrast to the heavy-tailed Pareto distribution in rogue waves, this mechanism can be used to compare with the real incident light intensity, rather than the same average intensity with output light. Thus the proposal demonstrates the potential of classifying and identifying light with different nature, and the most importantly, as well as generating light sources meeting various application requirements according to the presented rules. Most notably, the bunching and superbunching are distinguishable in super-Poissonian statistics using our mechanism.

We start to introduce the new mechanism on classification for antibunching, bunching or superbunching by Gamma fitting according to only one characteristic parameter $\alpha$. In general, the conventional approaches to discriminate antibunching, bunching or superbunching categories make use of calculating the normalized second-order coherence function ${g^{(2)}}(\tau )$ of light
\begin{eqnarray}
{g^{(2)}}(\tau ) = \frac{{\left\langle {\bar I(t)\bar I(t + \tau )} \right\rangle }}{{{{\left\langle {\bar I(t)} \right\rangle }^2}}},\
\label{eq:one}
\end{eqnarray}
where $\tau$ denotes the delay time, $\bar I(t)$ is the average of the intensity over a cycle of the oscillation and the brackets $\left\langle  \ldots  \right\rangle$ denote the statistical or longer-time average. For delay times $\tau$ much shorter than coherence time ${\tau _c}$, the degree of second-order coherence ${g^{(2)}}(0)$ is expressed in terms of the mean $\bar I$ and variance ${(\Delta I)^2}$ of light as \cite{Loudon1983The}
\begin{eqnarray}
{g^{(2)}}(0) = 1 + \frac{{{{(\Delta I)}^2}}}{{{{\bar I}^2}}}.\
\label{eq:two}
\end{eqnarray}

It turns out that the degree of second-order coherence depends on the statistical property of light, especially relationship between the mean and variance value of light. In most cases, the mean and variance value of statistical distributions are totally independent of each other. Interestingly enough, however, there is exactly a characteristic parameter ratio between the mean and variance value of some special distributions, such as Gamma, exponential, Poisson distribution, etc. It is well known that Poisson, chi-square, exponential distribution are special cases of general Gamma distribution, here we propose a new mechanism to discriminate different categories of antibunching, bunching and superbunching by Gamma fitting.
\begin{figure}[t]
\includegraphics[width=8.6cm]{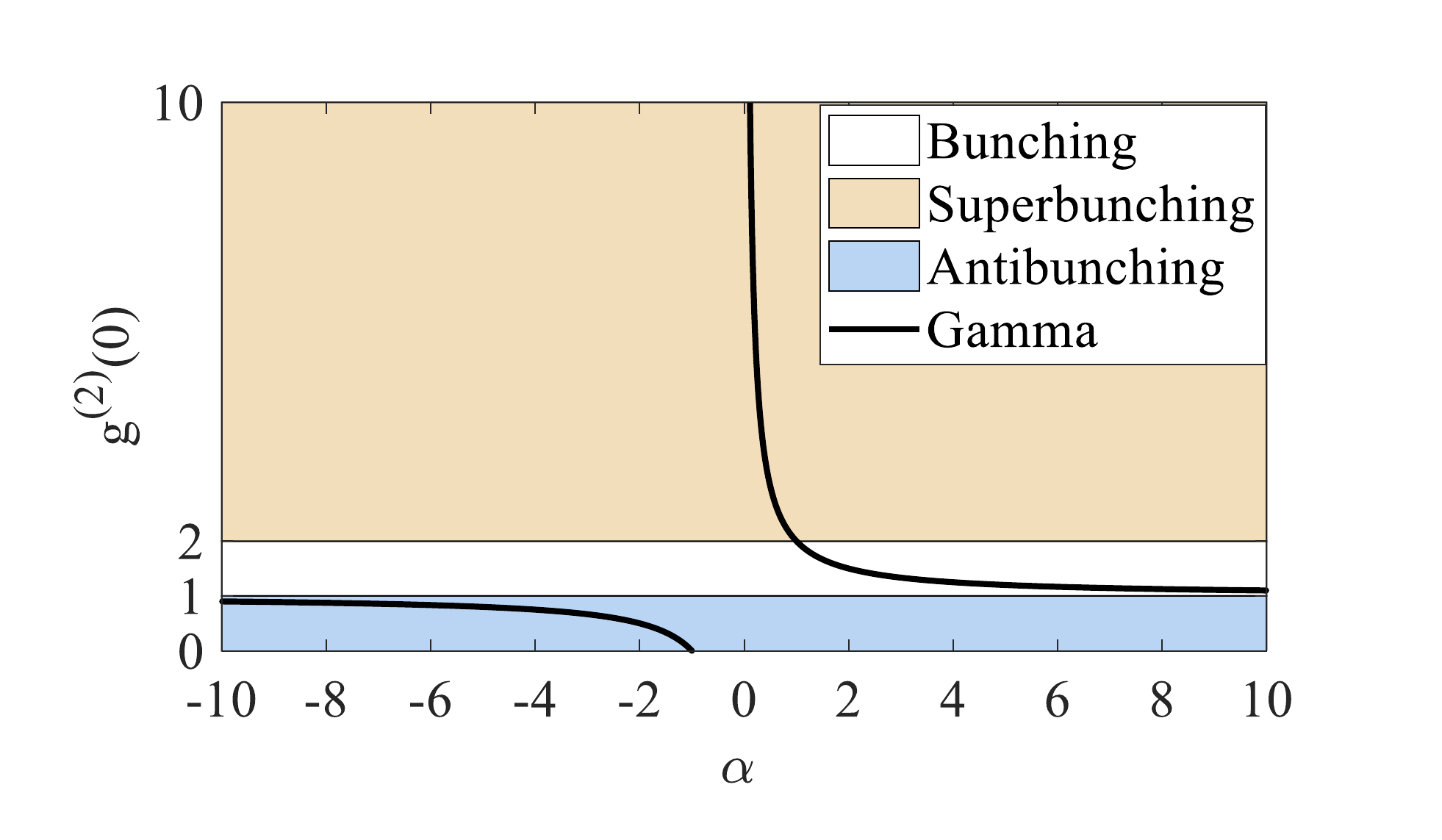}
\caption{\label{fig:1} The different regions of antibunching, bunching and superbunching for light obeying Gamma distribution.}
\end{figure}

As a preliminary to introduction of the proposed mechanism, a knowledge of statistical property of Gamma distribution is useful for understanding. The main focus of our model is on light intensity obeying the general Gamma statistical distribution, which probability density function is given by
\begin{eqnarray}
f\left( {I\left| {\alpha ,\beta } \right.} \right) = \frac{{{\beta ^\alpha }}}{{\Gamma \left(\alpha \right)}}{I^{\alpha  - 1}}{e^{ - \beta I}},\
\label{eq:three}
\end{eqnarray}
where $\Gamma ( \cdot )$ is the Gamma function, $\alpha$ is a shape parameter, and $\beta$ is an inverse scale parameter. The mean and variance value of Gamma distribution are $\bar I{\rm{ = }}\alpha {\rm{/}}\beta$ and ${(\Delta I)^2} = \alpha /{\beta ^2}$ respectively. It is noteworthy that the ratio of, square of mean to variance, is exactly the shape parameter $\alpha$, namely ${\bar I^2}/{(\Delta I)^2} = \alpha$. Therefore, it is concluded that the degree of second-order coherence ${g^{(2)}}(0)$ of light obeying Gamma distribution is determined by shape parameter $\alpha$.

With the attractive findings, the degree of second-order coherence of light obeying Gamma distribution can be simplified as
\begin{eqnarray}
{g^{(2)}}(0) = 1 + \frac{1}{\alpha }.\
\label{eq:four}
\end{eqnarray}
According to the obtained relationship, three different categories of antibunching, bunching and superbunching for light can be charted as
\begin{eqnarray}
\left\{\begin{array}{ll}
{{\alpha } < 0},&{antibunching}\\
{{\alpha } > 1},&{bunching}\\
{0 < {\alpha } < 1},&{superbunching.}
\end{array}\right.
\label{eq:five}
\end{eqnarray}

As shown in Fig.~\ref{fig:1}, the diagram is broadly divided into three regions, defined by antibunching ($\alpha  < 0$), bunching ($\alpha  > 1$) and superbunching ($0 < \alpha  < 1$). Additionally, it shows that the degree of second-order coherence ${g^{(2)}}(0)$ of light decreases as the shape parameter $\alpha$ increases in respective domains. Thus the proposal can be used to directly judge whether the light source is antibunching, bunching or superbunching by Gamma fitting according to only one parameter value $\alpha$, instead of conventional approach to calculating the ${g^{(2)}}(\tau )$ with light intensity. And what's more, the light source with various required properties can be generated in accordance with $\alpha$ value range.
\begin{figure}[t]
\includegraphics[height=5.44cm,width=8cm]{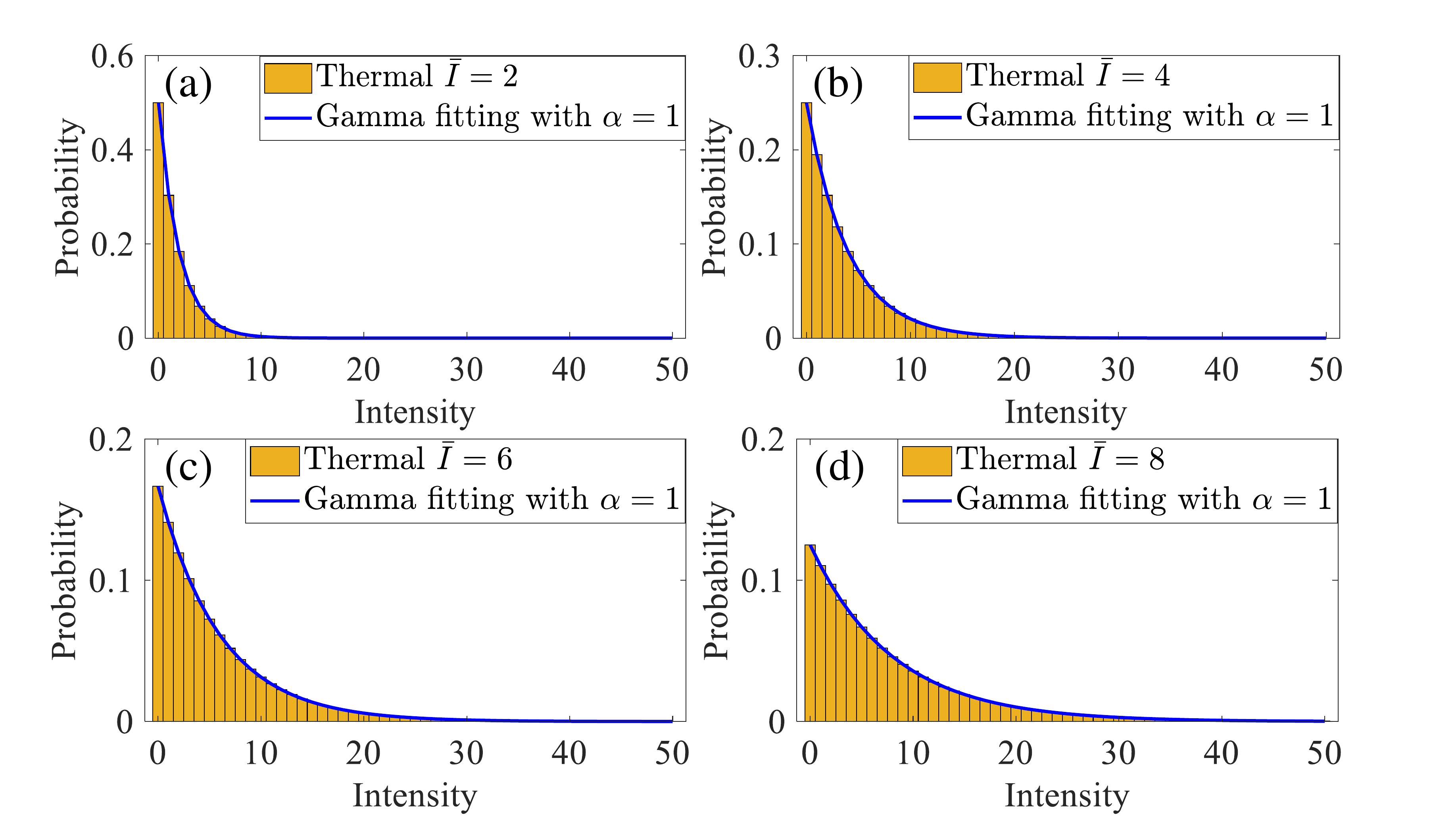}
\caption{\label{fig:2} Thermal light ($\alpha  = 1$) with different average intensities by Gamma fitting.}
\end{figure}

To explain and perform identification of light sources in the presented mechanism, the statistic properties of light sources are first discussed in detail. In most cases, the statistical distributions of classical light source can be characterized by general Gamma distribution, here thermal and coherent light are subjects of our models. The statistical distribution of thermal light is exponential distribution, which probability density function is given by
\begin{eqnarray}
f(I\left| \lambda  \right.) = \lambda {e^{ - \lambda I}}.\
\label{eq:six}
\end{eqnarray}
${\lambda > 0}$ denotes the rate parameter. The exponential distribution is a special case of Gamma distribution with the shape parameter $\alpha  = 1$ and the inverse scale parameter $\beta = \lambda$, the equivalent relation is
\begin{eqnarray}
Exp(\lambda ) = Gamma(1,\lambda ).\
\label{eq:seven}
\end{eqnarray}
It is evident from Eq.~(\ref{eq:seven}) that the exponential distributions with different $\lambda$, or the mean $\bar I{\rm{ = 1/}}\lambda$, maintain the constant shape parameter $\alpha = 1$ by Gamma fitting. According to the rules that thermal light obeying exponential distribution with different average intensities remain constant shape parameter $\alpha = 1$ by Gamma fitting, as shown in Fig.~\ref{fig:2}, one can distinguish thermal light from other light sources.

\begin{figure}[t]
\includegraphics[height=5.44cm,width=8cm]{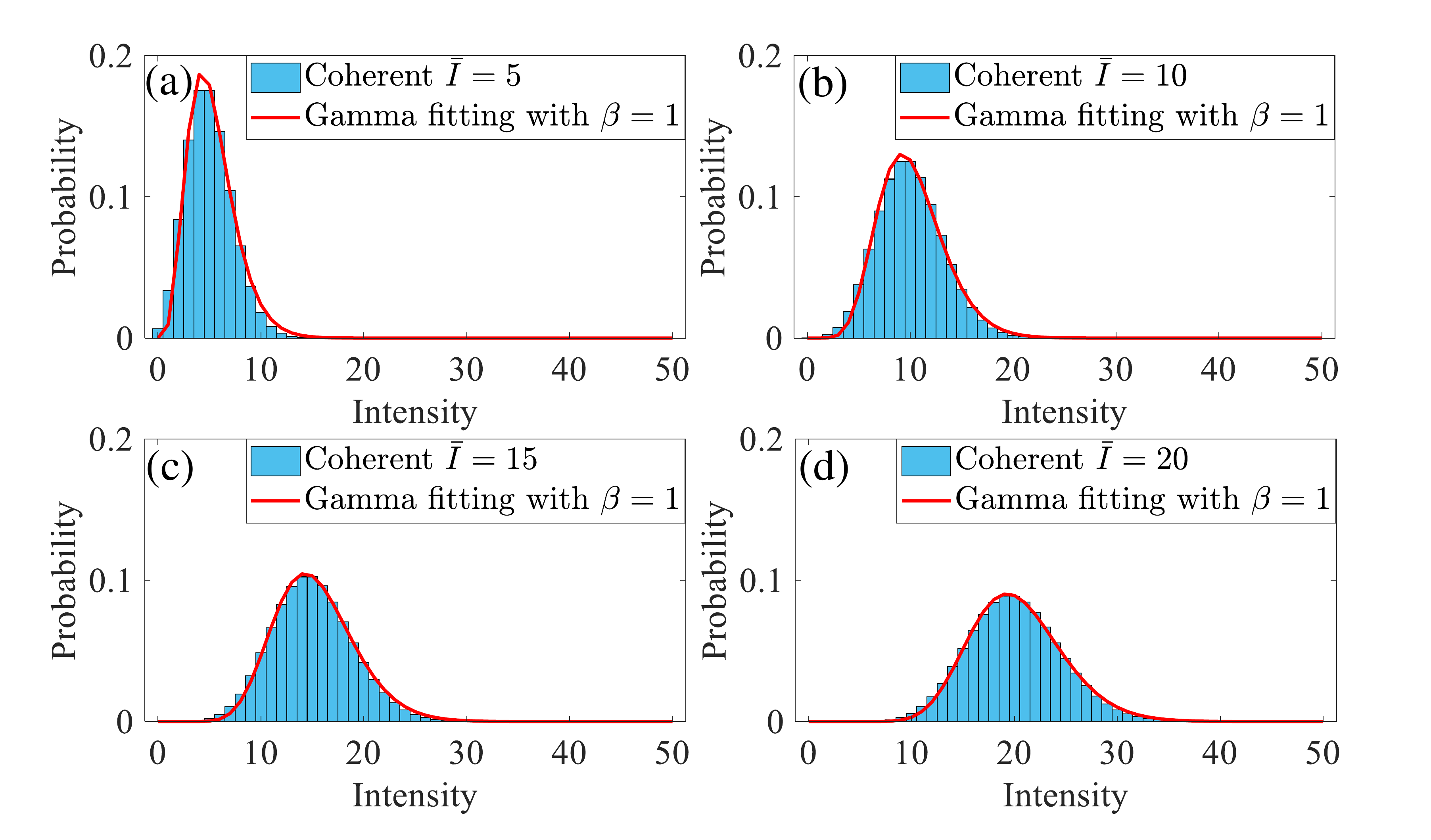}
\caption{\label{fig:3} Coherent light ($\beta = 1$) with different average intensities by Gamma fitting.}
\end{figure}

An alternative type of classical light source is coherent light, which statistical distribution is Poisson distribution, which probability density function is given by
\begin{eqnarray}
f(I\left| \lambda  \right.) = \frac{{{\lambda ^I}}}{{I!}}{e^{ - \lambda }}.\
\label{eq:eight}
\end{eqnarray}
${\lambda}$ denotes Poisson parameter. Suppose that the Gamma distribution is characteristic of the shape parameter $\alpha  = I + 1$ and the inverse scale parameter $\beta = 1$, which attains Gamma probability density function $f(x\left| {I + 1,1} \right.)$. As the variable $x = {\lambda}$, the function equals the probability density function of Poisson distribution
\begin{eqnarray}
Poisson(\lambda ) = Gamma(x = \lambda \left| {I + 1,1} \right.).\
\label{eq:nine}
\end{eqnarray}
Similarly, Eq.~(\ref{eq:nine}) shows that the Poisson distributions with different $\lambda$, or the mean $\bar I = \lambda$, maintain the inverse scale parameter $\beta = 1$ by Gamma fitting. According to the rules, we can determine coherent light obeying Poisson distribution differing from other light sources, as shown in Fig.~\ref{fig:3}. Therefore in classical optics, the proposed mechanism can be used to classify and generate typical light sources based on the presented mechanism.

\begin{figure*}[t]
\begin{minipage}[t]{0.7\textwidth}
\includegraphics[scale=0.38]{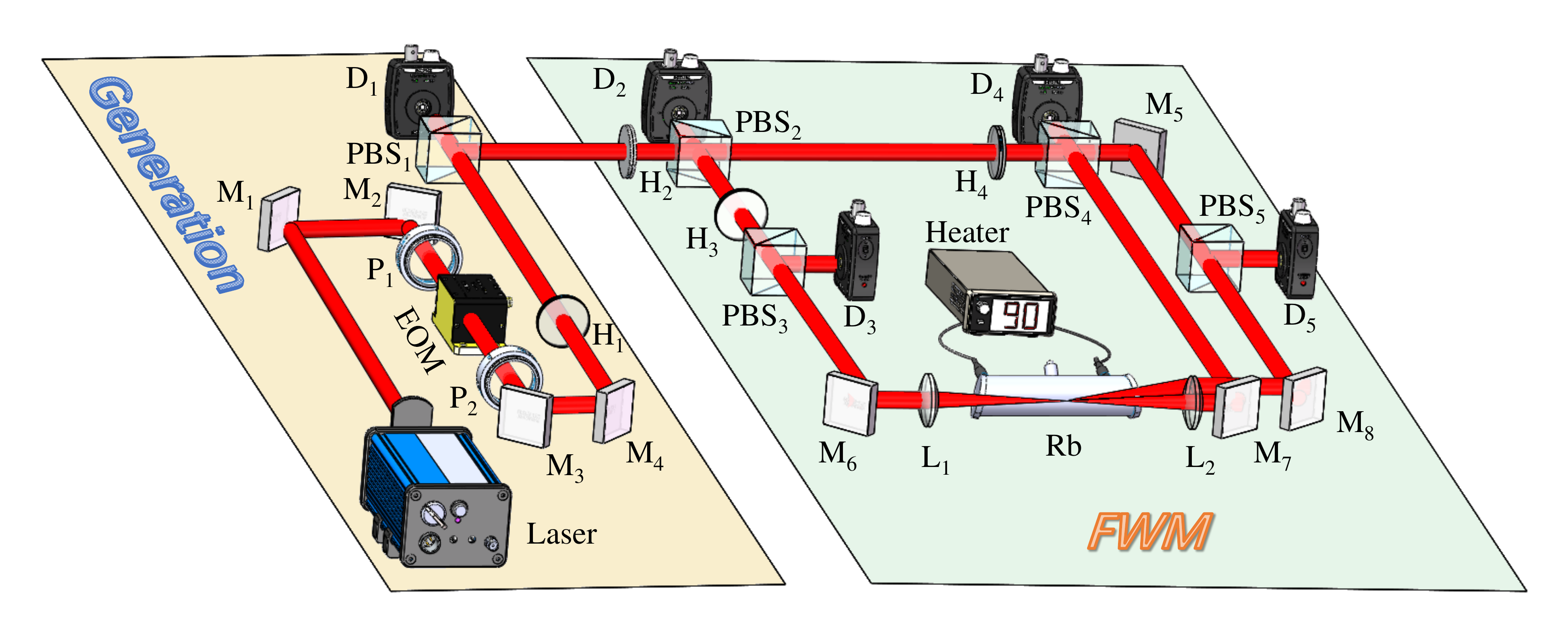}
\caption{\label{fig:4} Schematics of the experimental setup.}
\end{minipage}%
\begin{minipage}[t]{0.3\textwidth}
\centering
\includegraphics[scale=0.22]{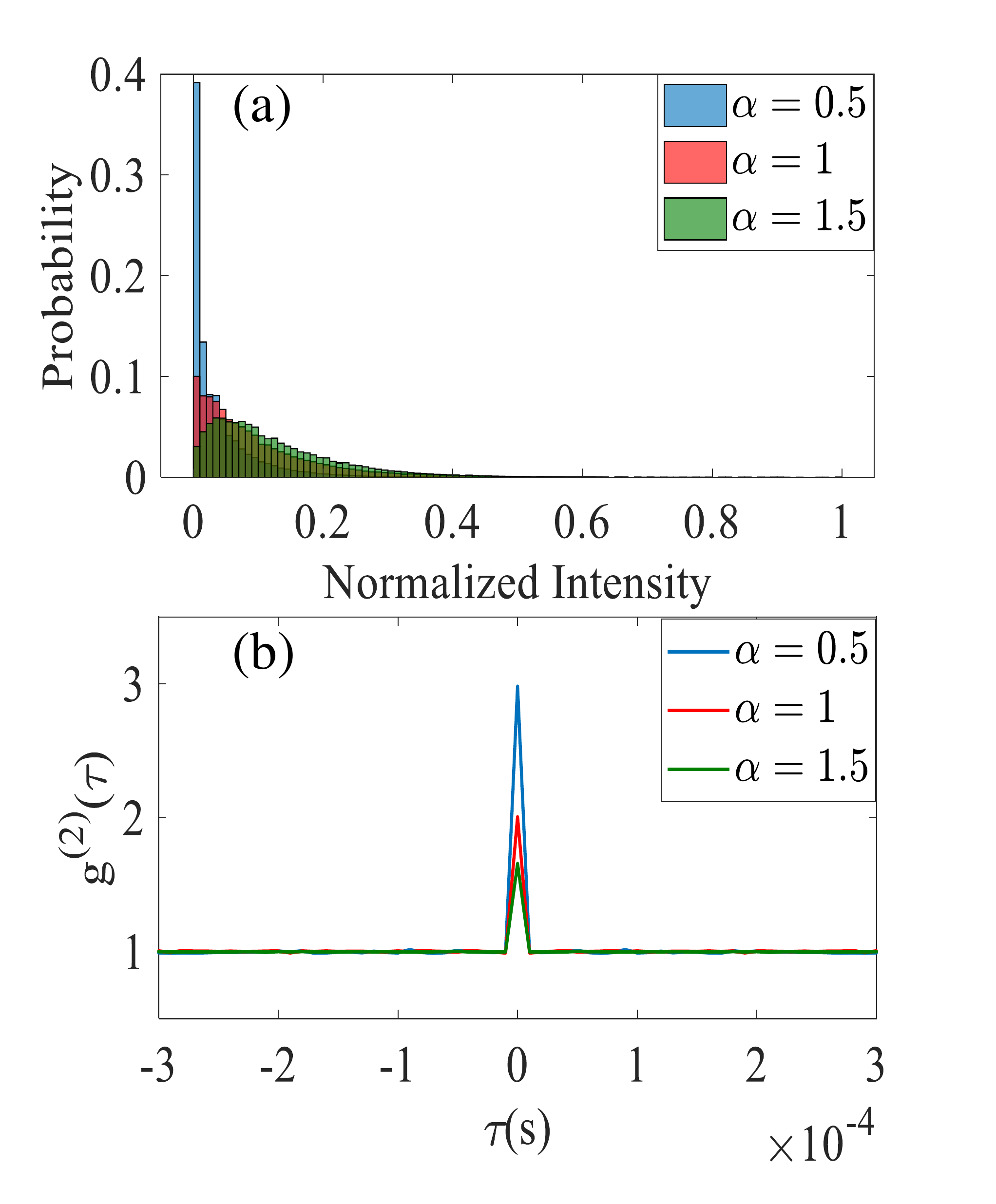}
\caption{\label{fig:5} (a) Statistical distributions and (b) ${g^{(2)}}(\tau )$ for different $\alpha$ values}
\end{minipage}
\end{figure*}

As per our earlier work of Ref. \cite{Zhang:20}, the superbunching effect was observed from four-wave mixing (FWM) process. To demonstrate the presented mechanism for classification and generation of light sources, we select FWM process as experimental verification, which experimental setup is shown in Fig.~\ref{fig:4}. The left part: A beam from a single-mode tunable diode laser with center wavelength 780.24 nm went through an electro-optical modulator, whose intensity can be manipulated into required light sources by regulating two linear polarizers P$_1$ and P$_2$ and inputting Gamma signals with corresponding $\alpha$ values in accordance with this mechanism. By adjusting a half-wave plate H$_1$ and polarization beam splitter PBS$_1$, the transmitted part of light, or the intensity distributions of light sources were measured by photoelectric detector D$_1$. Suppose that initially however that an ideal detector is available, with response time and sampling interval much shorter than the coherence time ${\tau _c}$. Specifically, the Gamma signals with $\alpha  = 0.5, 1, 1.5$ were inputted, then one generated the superbunching, thermal, bunching light, which statistical distributions and ${g^{(2)}}(\tau )$ equaling 3, 2, 1.6 are shown in Fig.~\ref{fig:5}.

Right part: One selected the generated thermal light as input light in FWM process. The reflected light from PBS$_1$ was divided into two beams by a set of half-wave plate H$_2$ and polarization beam splitter PBS$_2$. This reflected beam with S-polarization component was calibrated into P-polarization component via the second set of half-wave plate H$_3$ and polarization beam splitter PBS$_3$, named as forward pump beam. It propagated inside a rubidium atomic vapor. Meantime, the transmitted beam from PBS$_2$ was further divided into two parts by a set of half-wave plate H$_4$ and polarization beam splitter PBS$_4$. The reflected part with S-polarization served as the backward pump beam and counter-propagated with the forward pump beam for Doppler-free. The transmitted part through polarization beam splitter PBS$_5$ was chosen as the probe beam with P-polarization. Note that, they satisfied the phase-matching condition of a degenerated-FWM process. The two convex lenses L$_1$ and L$_2$ were used to focus three incident beams at a point in Rb cell. The Rb cell was manipulated by a temperature control heater with digital display, which can regulate and maintain the temperature of Rb vapor. The temperature of Rb cell was set to $90^\circ C$ in this experiment. The FWM signal was generated from the opposite direction of probe beam with S-polarization. High-speed photoelectric detectors D$_2$, D$_3$, D$_4$ and D$_5$ were employed to detect the intensity of probe, backward pump, forward pump, and FWM beam through nonlinear interaction, respectively. 
\begin{figure}[b]
\includegraphics[height=8cm,width=8.6cm]{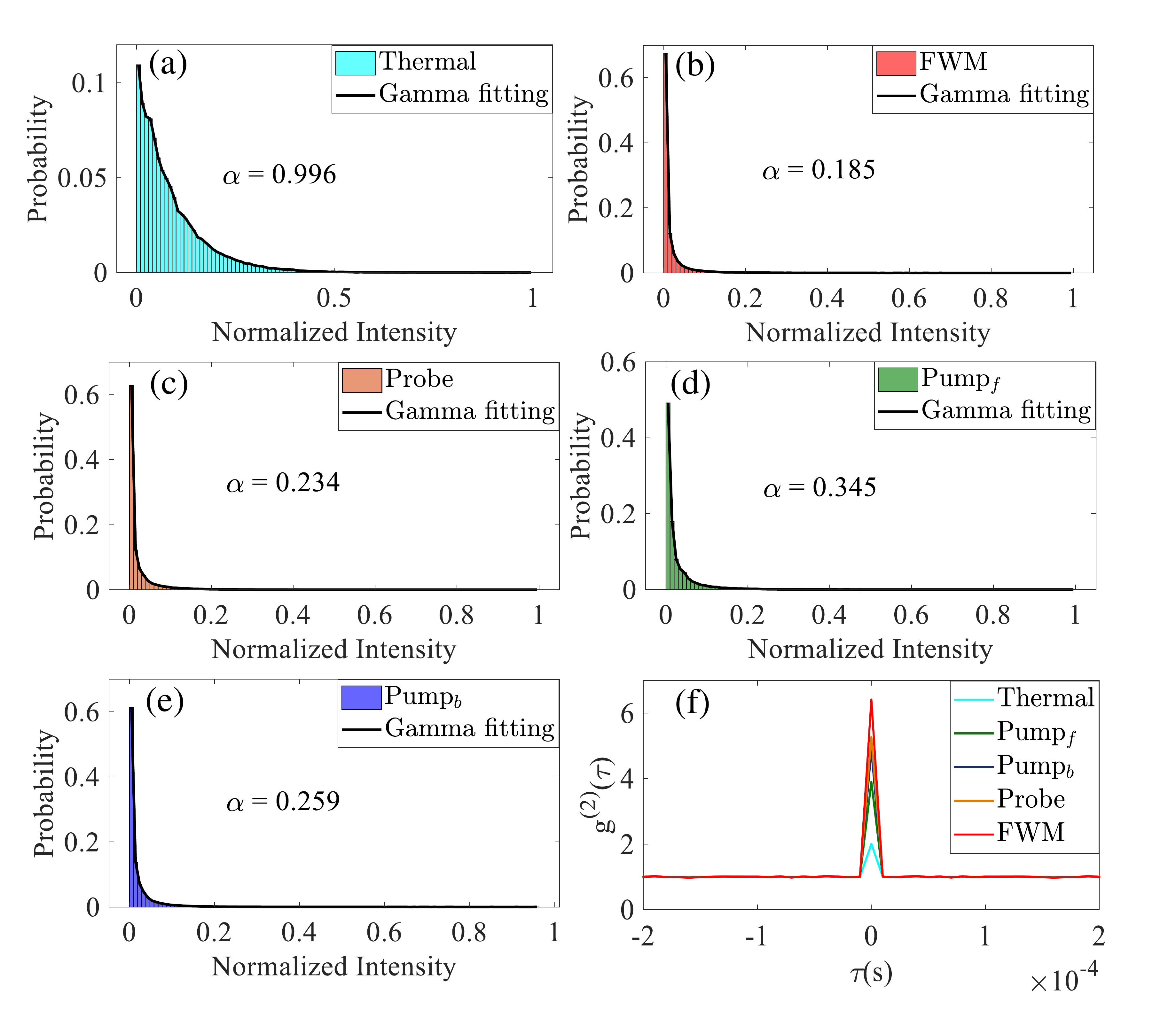}
\caption{\label{fig:6} (a)-(e) The statistical intensity distribution by Gamma fitting and (f) the second-order coherence function ${g^{(2)}}(\tau)$ of beams throughout FWM.}
\end{figure}

Figure~\ref{fig:6}(a)-(e) shows that the statistical intensity distributions of input beam and four output beams through FWM process are fitted with Gamma distribution. As seen in that, for the input beam, or thermal light, the shape parameter $\alpha = 0.996$ was observed, and the results showed the FWM, probe, forward and backward pump distribution with $\alpha = 0.185$, $0.234$, $0.345$ and $0.259$, respectively. Note that all the experimental results returned ${\rm{95\% }}$ confidence intervals for the parameter estimates. To verify the presented mechanism, the degree of second-order coherence ${g^{(2)}}(0)$ of beams were calculated by measured intensity throughout the FWM process, as shown in Fig.~\ref{fig:6}(f). It showed that all the FWM signal, probe, forward and backward pump beam achieved superbunching light with ${{\rm{g}}^{(2)}}(0) = 6.405$, $5.274$, $3.899$ and $4.861$. One can see that the obtained experimental results fully agreed with the presented theory relationship and classification rules, as Eqs.~(\ref{eq:four}) and (\ref{eq:five}).

\begin{figure}[t]
\includegraphics[width=8.6cm]{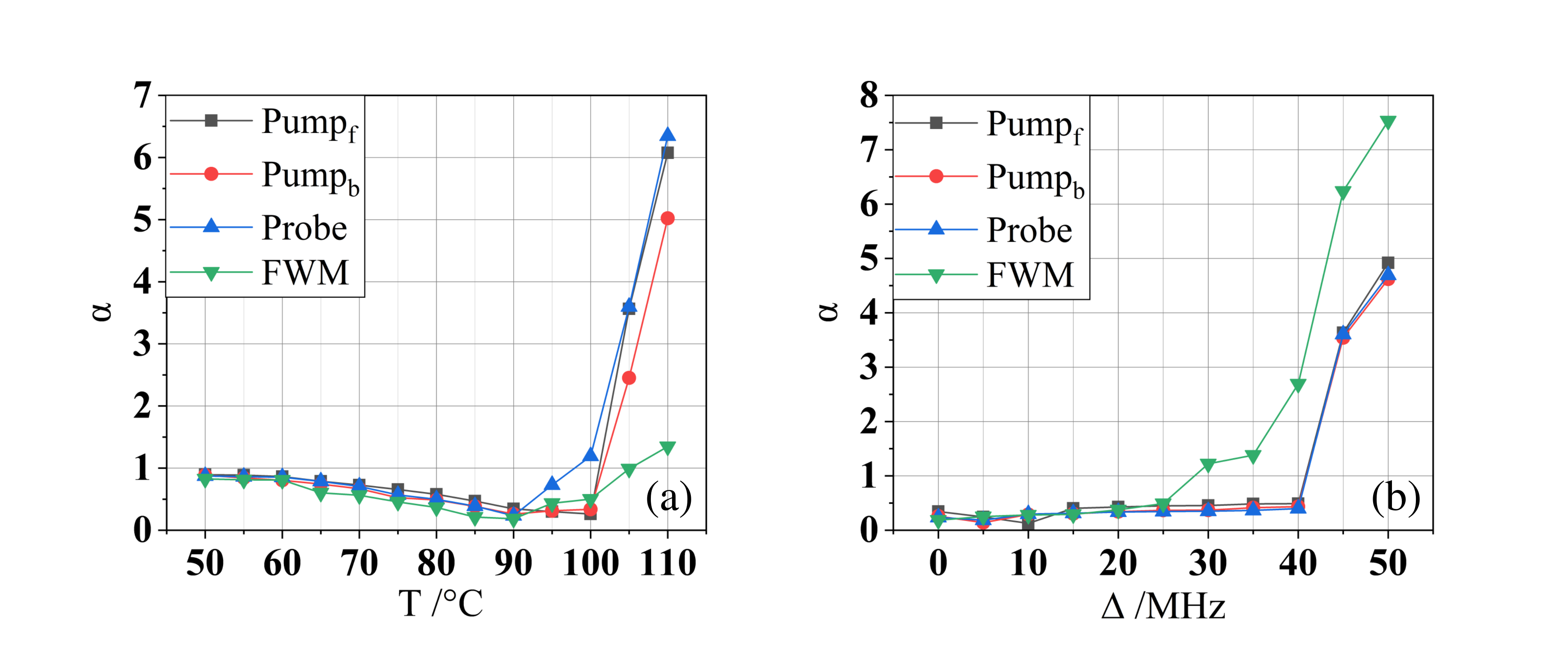}
\caption{\label{fig:7} Dependence of the measured shape parameter $\alpha$ of beams on (a) temperature $T$ and (b) laser detuning $\Delta$.}
\end{figure}

Next we further investigated how statistical intensity distribution of beams changed as influencing factors varied in nonlinear process, such as temperature $T$ and laser detuning $\Delta$. First the temperature varied from $50^\circ C$ to $110^\circ C$ by heater, and other experimental parameters were the same as before. The dependence of the measured shape parameter $\alpha$ by Gamma fitting of beams on temperature $T$ are obtained in Fig.~\ref{fig:7}(a). It was found that the shape parameter $\alpha$ of all beams decreased to the lowest values at the beginning and then rose sharply as temperature reached a certain value. As the degree of nonlinear interaction between incident beams and medium enhanced when temperature increased, the modulations for beams through process became stronger. When the saturation point of temperature was reached, nonlinear interaction weakened due to the presence of absorption effect \cite{Zhang2009Multi}.

Then the influence of laser detuning $\Delta$ on the measured shape parameter $\alpha$ of beams are shown in Fig.~\ref{fig:7}(b). The laser detuning varied from resonant to 50 MHz by tuning laser, and other experimental conditions were the same as before. As shown in the figure, the measured shape parameter $\alpha$ of FWM signal ascended gradually as laser detuning increased, while the other three beams were basically stable and then increased at some point. Since the FWM frequency spectrum was relatively narrower than the other three beams, which moved more rapidly away from the strong interaction area.

\begin{figure}[t]
\includegraphics[width=8.6cm]{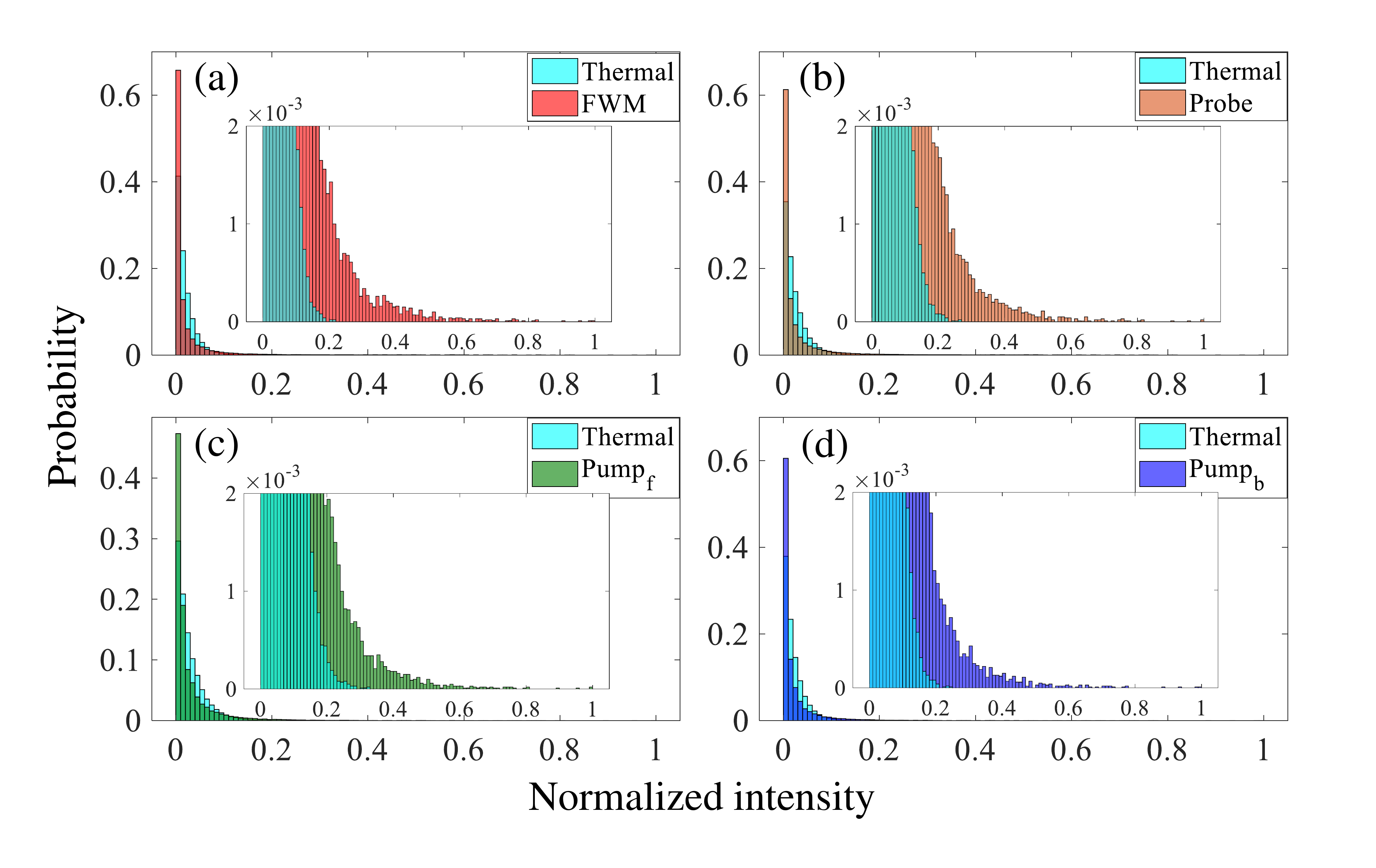}
\caption{\label{fig:8} The measured heavy-tailed distribution of (a) FWM signal, (b) probe beam, (c) forward and (d) backward pump beams, compared to the thermal distributions with the same mean value.}
\end{figure}

By contrast, a review of the literature \cite{manceau2019indefinite} undertaken found that the an extreme heavy-tailed distribution is the Pareto one with indefinite-mean value at Pareto index $k < 1$. To test whether our experimental results is also extreme heavy-tailed distribution, the obtained light intensity distribution of beams were fitting by Pareto distribution, as shown in Fig.~\ref{fig:8}. It was concluded that the all obtained beams in this experiment were extreme heavy-tailed distribution with Pareto index $k = 0$.

However, the “heavy tail” behavior of light, after squeezing and optical harmonics in \cite{manceau2019indefinite}, must be illustrated by comparing to the incident statistical distribution with the same average intensity of output light, rather than real input light, then the superbunching effect was demonstrated by calculating second-order correlation function. In contrast, the output light can be compared with the incident distribution with the real input light intensity in the presented mechanism, even the beams with the different average intensity can be directly classified and determined according to one characteristic parameter $\alpha$ using this mechanism, rather than calculating second-order correlation function for criterion.

In summary, the presented mechanism can be used to directly classify and generate antibunching, bunching or superbunching categories of light by Gamma fitting according to only shape parameter $\alpha$, instead of conventional approach to calculating the second-order coherence function ${g^{(2)}}(\tau )$ of light intensity. Most notably, the bunching and superbunching are distinguishable in super-Poissonian statistics using our mechanism. For classical light sources, the identification and generation of thermal and coherent light have been performed by Gamma fitting in terms of shape parameter $\alpha$ or inverse scale parameter $\beta$. Experimental verification of beams from FWM process is in agreement with the proposal, and the influence of temperature $T$ and laser detuning $\Delta$ on the measured results have been investigated. In contrast to the heavy-tailed Pareto distribution in rogue waves, this mechanism can be used to compare with the real incident light intensity, rather than the same average intensity with output light. The proposal demonstrates the potential of classifying and identifying light with different nature, and the most importantly, as well as generating light sources meeting various application requirements according to the presented rules. 

This work is supported by Shaanxi Key Research and Development Project (Grant No. 2019ZDLGY09-10); Key Innovation Team of Shaanxi Province (Grant No. 2018TD-024); 111 Project of China (Grant No. B14040).

\bibliographystyle{apsrev4-2}
\bibliography{apssamp}

\end{document}